\documentclass{article}
\usepackage{spconf,amsmath,graphicx}

\usepackage{enumitem}

\usepackage{xcolor}

\newcommand{\eg}{{e.g.,}}
\newcommand{\etal}{{et~al.}}

\title{{SEP-28k}: A Dataset for Stuttering Event Detection\\ from Podcasts with People Who Stutter}
\name{Colin Lea*, Vikramjit Mitra*, Aparna Joshi, Sachin Kajarekar, Jeffrey P. Bigham}
\address{Apple}

\begin{document}

\maketitle

\begin{abstract}
The ability to automatically detect stuttering events in speech could help speech pathologists track an individual's fluency over time or help improve speech recognition systems for people with atypical speech patterns. 
Despite increasing interest in this area, existing public datasets are too small to build generalizable dysfluency detection systems and lack sufficient annotations. 
In this work, we introduce Stuttering Events in Podcasts (SEP-28k), a dataset containing over 28k clips labeled with five event types including blocks, prolongations, sound repetitions, word repetitions, and interjections. 
Audio comes from public podcasts largely consisting of people who stutter interviewing other people who stutter. 
We benchmark a set of acoustic models on SEP-28k and the public FluencyBank dataset and highlight how simply increasing the amount of training data improves relative detection performance by 28\% and 24\% F1 on each. 
Annotations from over 32k clips across both datasets will be publicly released. 
\end{abstract}


\section{Introduction}
\label{sec:intro}

Dysfluencies in speech such as sound repetitions, word repetitions, and blocks are common amongst everyone and are especially prevalent in people who stutter.
Frequent occurrences can make social interactions challenging and limit an individual's ability to communicate with ubiquitous speech technology including Alexa, Siri, and Cortana~\cite{BrewerCSCW18,ClarkCUI20,USAToday,Moolya,Slate}. 
In this work we investigate the ability to automatically detect dysfluencies, which may be valuable for clinical assessment or development of accessible speech recognition technology.

\begin{figure}[t!]
\centering
\includegraphics[width=0.39 \paperwidth]{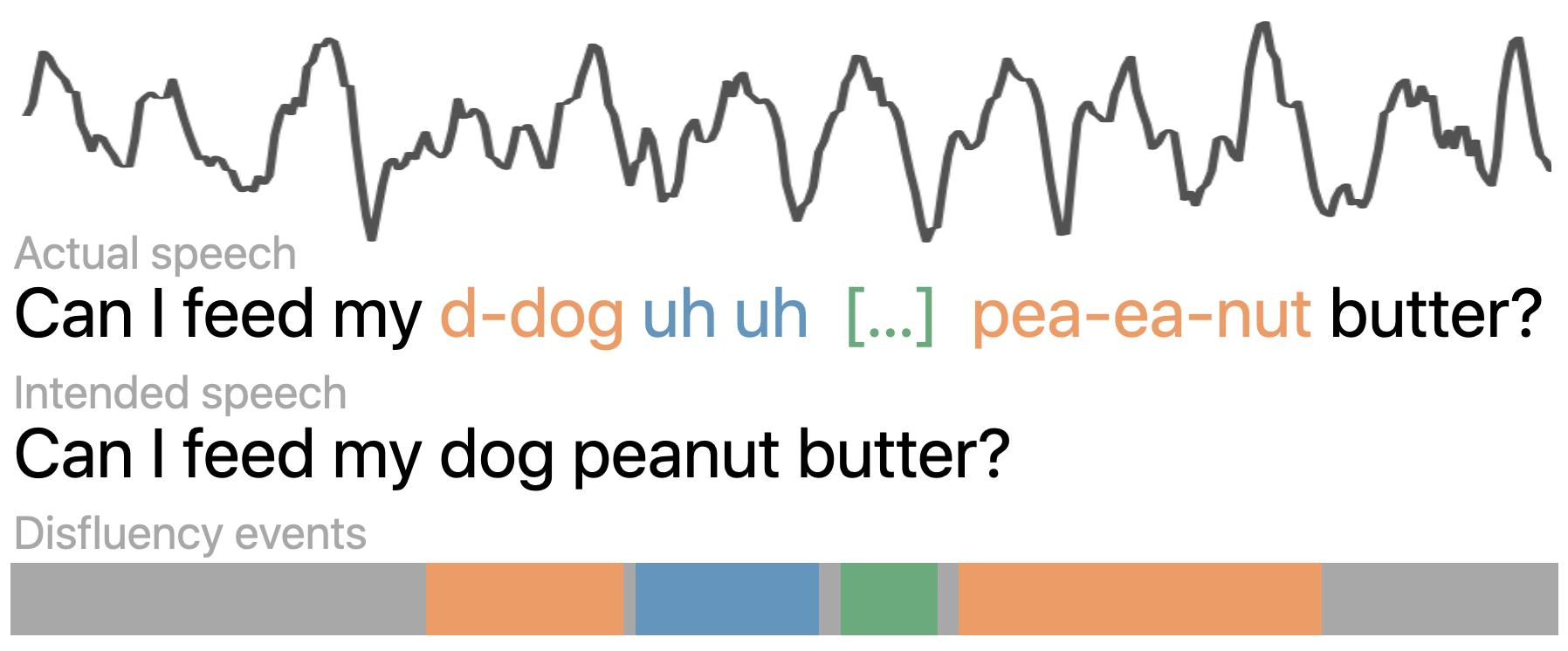}
\caption{Speech from someone who stutters may contain events including sound repetitions (orange), interjections (blue), blocks/pauses (green), or other events that make speech recognition challenging.} 
\label{fig:problem}
\end{figure}

This problem is challenging because there are many variations in how a given individual expresses each dysfluency type, in the patterns of dysfluencies between users, and even how the situation or environment affects their speech.
For example, an individual may stutter when conversing but not while reading aloud; when talking with a teacher but not a friend; or when stressed before an exam but not in every day-to-day interaction. 
The speech pathology community has spent decades characterizing, developing diagnosis tools, and developing strategies to mitigate these behaviors~\cite{VanRiper1982,SSI,valente2015event,ingham1993time}, however, there has been limited success in taking these learnings and applying them to speech recognition technology, where individuals may be frequently cut off or have their speech inaccurately transcribed.
%

A major bottleneck in this area is that dysfluency datasets tend to be small and have few or inconsistent annotations not inherently designed for work on speech recognition tasks. 
Kourkounakis \etal~\cite{kourkounakis2020detecting} used 800 speech clips (53 minutes) with custom annotations to detect dysfluencies from 25 children who stutter using the UCLASS dataset~\cite{UCLASS}. 
Riad \etal~\cite{riad2020identification} performed a similar task using 1429 utterances from 22 adults who stutter with the recent FluencyBank~\cite{FluencyBank} dataset. 
Bayer \etal~\cite{BayerlTSD2020} collected a 3.5 hour German dataset with 37 speakers and developed a model for automated stuttering severity assessment.
Unfortunately, none of the annotations from these efforts have been released. 
A core contribution of our paper is the introduction of the Stuttering Events in Podcasts dataset (SEP-28k)
dataset which contains 28k annotated clips (23 hours) of speech curated from public podcasts.
We have released these along with annotations for 4k clips (3.5 hours) from FluencyBank targeted at stuttering event detection.

\begin{table*}[t!]
    \centering
\begin{tabular}{|l|l|c|c|c|}
    \hline
    \textbf{Stuttering Labels} & \textbf{Definition} & \textbf{SEP-28k} & \textbf{FluencyBank}  \\ 
    \hline
    Block & Gasps for air or stuttered pauses & 12.0\% & 10.3\%   \\
    Prolongation & Elongated syllable \textit{``M[mmm]ommy''} & 10.0\% & 8.1\%   \\
    Sound Repetition & Repeated syllables \textit{``I [pr-pr-pr-]prepared dinner''} &  8.3\% & 13.3\%   \\
    Word/Phrase Repetition &  \textit{``I made [made] dinner''}  & 9.8\% & 10.4\%   \\
    No dysfluencies & Affirmation that there are no discernable dysfluencies & 56.9\% & 54.1\%   \\
    Interjection & Filler words e.g., \textit{``um,'' ``uh,'' \& ``you know''} & 21.2\% & 27.3\%   \\
    \hline
    \textbf{Non-dysfluent Labels} & & &  \\
    \hline
    Natural pause & A pause in speech (not as part of a stutter event) & 8.5\% & 2.7\%  \\
    Unintelligible & It is difficult to understand the speech & 3.7\% & 3.0\%  \\
    Unsure & An annotator was unsure of their response & 0.1\% & 0.4\%  \\    
    No Speech & The clip is silent or only contains background noise & 1.1\% & -  \\
    Poor Audio Quality & There are microphone or other quality issues & 2.1\% & -  \\    
    Music & Music is playing in the background & 1.1\% & -  \\
    \hline
\end{tabular}
    \caption{Distribution of annotations in each dataset where at least two of three annotators applied a given label.}
    \label{tab:datainfo}
\end{table*}

The focus of this paper is on detection of five stuttering event types: Blocks, Prolongations, Sound Repetitions, Word/Phrase Repetitions, and Interjections.
Existing work has explored this problem using traditional signal processing techniques~\cite{Tripathi2018,Tripathi2020,Czyzewski2003}, language modeling (LM)~\cite{riad2020identification,alharbi2018lightly,HeemanInterSpeech16,AlharbiWOCCI17,Mahesha2016}, and acoustic modeling (AM)~\cite{Mahesha2016,kourkounakis2020detecting}. 
Each approach has be shown to be effective at identifying one or two event types typically on data from a small number of users. 
\underline{Prolongations}, or extended sounds, have been detected using short-window autocorrelations~\cite{Tripathi2020} and low-level acoustic models~\cite{kourkounakis2020detecting}. 
\underline{Word/phrase repetitions}, if they are well articulated, are easily detected using LM-based approaches~\cite{HeemanInterSpeech16}, with the caveat that single-syllable words such as in the phrase ``I-I-I am'' will often be smoothed into ``I am'' due to the underlying acoustic model and phrases like ``I am [am]'' may be pruned because the LM has never seen the word ``am'' repeated before. This is fine for speech recognition but bad for stuttering event analysis. 
Arjun \etal~\cite{Tripathi2020} addressed this repetition problem by segmenting pairs of subsequent words and analyzing correlations in their spectral features. 
\underline{Interjections}, including ``um'', ``uh'', ``you know'' and other filler words, are perhaps the easiest type to recognize with a language model if well articulated.
\underline{Blocks}, or gasps/pauses typically within or between words, are difficult to detect because the gasp for breath or pause is often inaudible. 
\underline{Sound repetitions} are also challenging because syllables may vary in duration, count, style, and articulation (\eg ``[moh-muh-mm]-ommy'').

Efforts in HCI have sought out an understanding of speech recognition needs for users with speech impairments, which is critical for framing problems like ours~\cite{BrewerCSCW18,ClarkCUI20,kane2020sense}.
\section{Data}
\label{sec:data}



\subsection{Stuttering Events in Podcasts (SEP-28k)}

We manually curated a set of podcasts, many of which contain speech from people who stutter talking with other people who stutter, using a two step process.
Shows were initial selected by searching metadata from a podcast search engine with terms related to dysfluencies such as \textit{stutter}, \textit{speech disorder}, and \textit{stammer}. 
This resulted in approximately 40 shows and 100s of hours of audio. 
Many of these were about speech disorders but did not contain high rates of speech from people who stutter. 
After culling down the data we extracted clips from 385 episodes across 8 shows. Specific show names and links to each episode can be found in the dataset respository. 

We extracted 40$-$250 segments per episode for a total of 28,177 clips.
Dysfluency events are more likely to occur soon before, during, or after a pause so we used a voice activity detector to extract 3-second intervals near pauses.
We varied where we sampled each interval with respect to a breakpoint to capture a more representative set of dysfluencies. 

\subsection{FluencyBank}
We used all of the FluencyBank~\cite{FluencyBank} interview data which contains recordings from 32 adults who stutter. 
As with Riad \etal~\cite{riad2020identification} we found the temporal alignment for some transcriptions and dysfluency annotations provided were inaccurate, so we ignored these and used the same process as SEP-28k to annotate 4,144 clips (3.5 hours).


\subsection{Annotations}
Annotating stuttering data is difficult because of ambiguity in what constitutes stuttering for a given individual. 
Repetitions, for example, can occur during stuttering events or when an individual wants to emphasize a word or phrase.
Speech may be unintelligible which makes it challenging to identify how a word was stuttered. 
We annotated our data using a variant of time-interval based assessment~\cite{valente2015event} in which audio recordings are broken into 3 second clips and annotated with binary labels as defined in Table~\ref{tab:datainfo}.
A clip may contain multiple stuttering event types along with non-dysfluency labels such as \textit{natural pause} and \textit{unintelligible speech}.
SEP-28k was also annotated with: \textit{no speech}, \textit{poor audio quality}, and \textit{music} to identify issues specific to this medium.

Clips were annotated by at least three people who received training via written descriptions, examples, and audio clips on how to best identify each dysfluency but were not clinicians. 
We measured Fleiss Kappa inter-annotator agreement and found word repetitions, interjections, sound repetitions, and no dysfluencies were more consistent (0.62, 0.57, 0.40, 0.39) and blocks and prolongations had only fair or slight agreement (0.25, 0.11). 
Blocks can be difficult to assess from audio alone; clinicians often rely on physical signs of grasping for air when making this assessment.
As such, results when using the block labels should be more speculative.

\subsection{Evaluation \& Metrics}
\label{ssec:metrics}
We use F1 score and Equal Error Rate to evaluate dysfluency detection where each annotation constitutes a binary label.
F1 is the harmonic mean of precision ($P$) and recall ($R$): $F1 = 2 \frac{P \cdot R}{P+R}$. Equal Error Rate (EER) is the point in the Receiver Operating Characteristic (ROC) curve where the false acceptance rate is equal to the false rejection rate and reflects how well the two classes are separated. 
The lower the EER, the better the performance of the model.
We report results for each label individually and as a combined ``Any'' label which includes all five stutter types.

SEP-28k is partitioned into three splits containing 25k samples for training, 2k for validation, and 1k for testing. 
FluencyBank is partitioned across the 32 individuals in the dataset: 26 individuals ($\sim$3.6k clips) for training, 3 ($\sim$500 clips) for validation, and 3 ($\sim$500 clips) for testing. 
We encourage others to explore alternative splits to tease out differences between speakers, podcasts, or other analyses.

\begin{figure}[t!]
\centering
\includegraphics[width=0.4 \paperwidth]{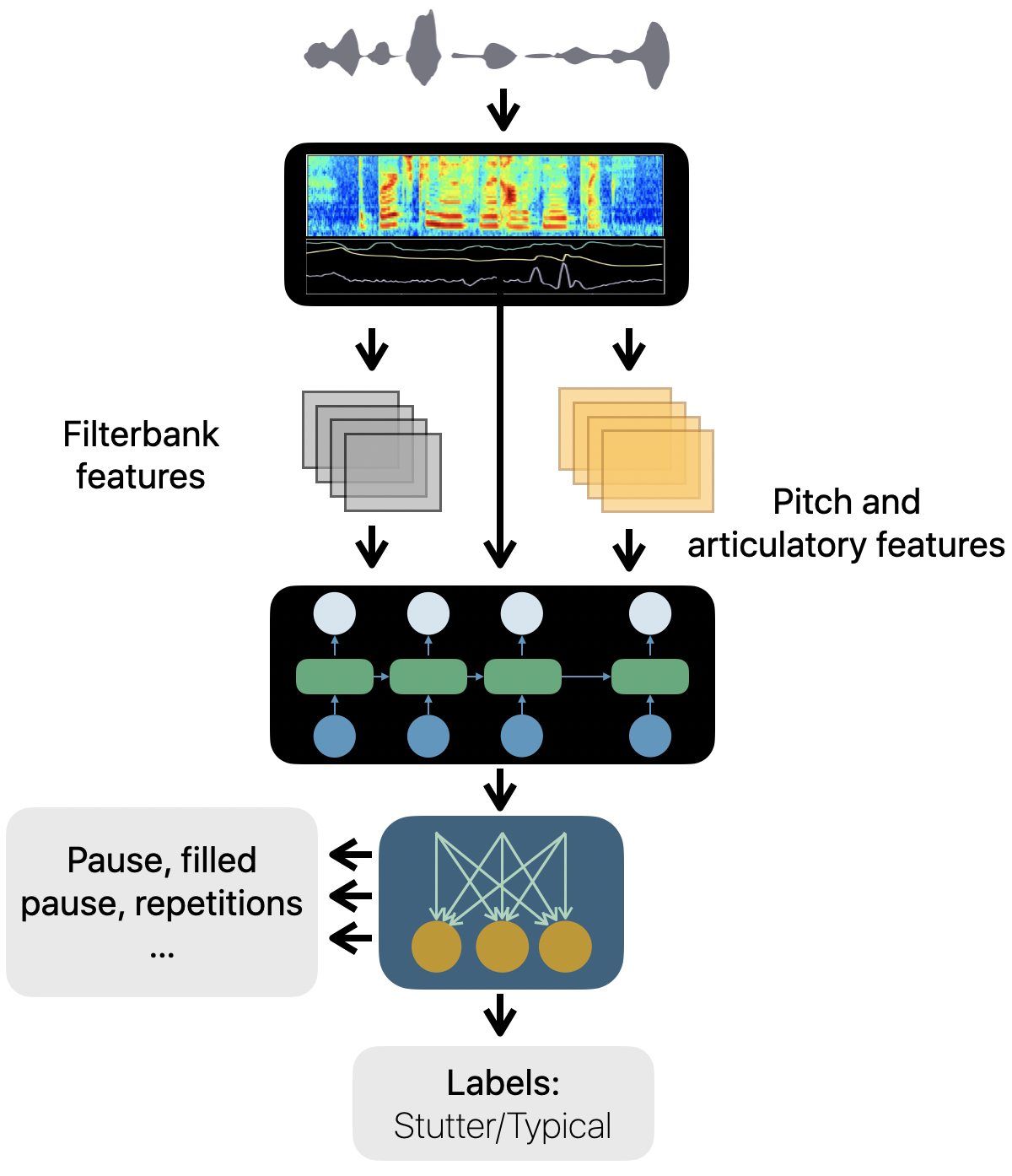}
\caption{Multi-feature acoustic stutter detection model}
\label{fig:revised_baseline}
\end{figure}

\section{Methods}
\label{sec:approach}



Our approach takes an audio clip, extracts acoustic features per-frame, applies a temporal model, and outputs a single set of clip-level dysfluency labels. 
We investigated baselines that are inspired by the dysfluency model in \cite{kourkounakis2020detecting} and alternative input features, model architectures, and loss functions. 


\subsection{Acoustic Features}
Our baseline input is a set of 40 dimensional mel-filterbank energy features ($M_{FB}$). We use frequency cut-offs at 0 $hz$ and 8000 $hz$, a 25 $ms$ window, and a sample rate of 100 $hz$. 
We compare with three additional feature types:
\begin{itemize}[noitemsep]
    \item $F_0$ (3 dim): pitch, pitch-delta and voicing features;
    \item $A_{TV}$ (8 dim): articulatory features in the form of vocal-tract ($TV$) constriction variables  \cite{mitra2017hybrid}. These define degree and location of constriction actions within the human vocal tract \cite{mitra2017hybrid, mitra2010retrieving} as implemented in~\cite{mitra2019interspeech};
    \item $F_{Phone}$ (41 dim): phoneme probabilities extracted from an acoustic model trained on LibriSpeech~\cite{librispeech} using a Time-depth Separable CNN architecture~\cite{timedepthseparable}.
\end{itemize}
Pitch, voicing, and articulatory features encode voice quality and often change across dysfluency events. We hypothesize these may improve detection of blocks or gasps. 
Phoneme probabilities may make it easier to identify sound repetitions where the same phoneme fires multiple times in a row.



\subsection{Model Architectures}
The baseline stutter detection model consists of a single-layer LSTM network and an improved model adds convolutional layers per-feature type and learns how the features should be weighted, as shown in Figure~\ref{fig:revised_baseline}. 
We refer to the latter as ConvLSTM. 
Feature maps from the convolution layer are combined after batch normalization and fed to the LSTM layer.
The temporal convolution size for $M_{FB}$ feature was set to 3 frames and for the remaining features were set to 5 frames.
We use unidirectional recurrent networks where the final state is fed into the per-clip classifier. 
Both models have two output branches: a fluent/dysfluent prediction and a soft prediction for each of the five event types. 




\subsection{Loss functions}
The baseline model has a single cross-entropy loss term. 
Our improved models are trained with a multi-task objective where the fluent/dysfluent branch has a weighted cross-entropy term with focal loss \cite{focalloss} and the per-dysfluency branch has a concordance correlation coefficient ($CCC$) loss using the inter-annotator agreement for each clip. 

Models were trained with a mini-batch size of 256, using the Adam optimizer, with an initial learning rate of 0.01. 
Early stopping was used based on cross-validation error. 
Networks had 64 neurons in recurrent and embedding layers. 




\section{Experiments \& Analysis}
\label{sec:experiments}


\subsection{Model Design}
Table~\ref{tab:table1} compares performance across features and architectures types.
Spectral features with pitch generally perform well and when using the improved model achieve best performance when adding articulatory signals. 
This improvement matches our intuition that variation in intonation and articulation coincides with dysfluent speech. 
The phoneme-based models perform worst, despite their ability to extract features one might think would be useful for sound repetitions.
The ConvLSTM and CCC loss moderately improve F1, likely because this loss explicitly encodes uncertainty in annotators.

Table~\ref{tab:per_dysfluency} shows performance per-dysfluency type. 
Performance is worse for Blocks and Word Repetitions.
These dysfluencies tend to last longer in time and have more variation in expression, which may contribute to the lower performance. 
Interjections and prolongations tend to have less variability and are easier to detect.
 SEP-28k performance is consistently worse than FluencyBank, likely given the larger variety of individuals and speaking styles. 


\begin{table}[]
\centering
\caption{ Weighted Accuracy (WA), F1-score and Equal Errors Rate ($EER$) from each model on FluencyBank (eval).}
\vspace{1mm}
\label{tab:table1}
  \begin{tabular}{lccc}
     & {$WA \uparrow$} & {$F1\uparrow$} & {$EER\downarrow$} \\
    \hline     
     \textbf{Baseline} (LSTM, XEnt) & \\
    $F_{Phone}$ & 74.6 & 74.8  & 24.7     \\     
    $M_{FB}$ & 77.7 & 75.8 & 23.8  \\
    $M_{FB}+F_0$ & 81.6 & 81.8 & 18.0  \\
    $M_{FB}+F_0+A_{TV}$ & 81.8 & 80.1 & 19.0
    \\
    \hline
    \textbf{Improved} (ConvLSTM, CCC) & \\
    $F_{Phone}$ & 80.8  & 80.2 & 17.1     \\        
    $M_{FB}$ & 83.0 & 81.9 & 16.1   \\
    $M_{FB}+F_0$ & 83.4 & 82.7 & 16.9  \\
    $M_{FB}+F_0+A_{TV}$ & 83.6 & 83.6 &  16.9 \\
    \hline
  \end{tabular}
\end{table}

\begin{table}[t]
    \caption{F1 score per dysfluency type with a baseline LSTM model (XEnt loss) trained using single- or multi-task learning (STL, MTL) and the Improved ConvLSTM model (CCC loss). Bl=Block, Pro=Prolongation, Snd=Sound Repetition, Wd=Word Repetition, Int=Interjection}
    \centering
\begin{tabular}{|l|c|c|c|c|c|c|}
    \hline
    \textbf{SEP-28k} & \textbf{Bl} & \textbf{Pro} & \textbf{Snd} & \textbf{Wd} & \textbf{Int} & \textbf{Any}  \\
    \hline
    Random & 13.7 & 12.8 & 9.5 & 4.3 & 13.6 & 46.0  \\
    Baseline (STL) & 54.9 & 65.4 & 57.2 & 60.7 & 64.9 & 61.5  \\
    Baseline (MTL) & 56.4 & 65.1 & 60.5 & 56.2 & 69.5 & 64.5 \\
    Improved & 55.9 & 68.5 & 63.2 & 60.4 & 71.3 & 66.8 \\
    \hline
    \textbf{FluencyBank} & \textbf{Bl} & \textbf{Pro} & \textbf{Snd} & \textbf{Wd} & \textbf{Int} & \textbf{Any}  \\
    \hline
    Random & 12.9 & 10.7 & 28.2 & 10.3 & 31.7 & 31.7 \\
    Baseline (STL) & 58.6 & 63.2 & 60.8 & 61.8 & 57.2 & 73.2 \\
    Baseline (MTL) & 54.6 & 67.6 & 74.2 & 55.8 & 75.0 & 74.8 \\
    Improved & 56.8 & 67.9 & 74.3 & 59.3 & 82.6 & 80.8 \\
    \hline    
\end{tabular}
    \label{tab:per_dysfluency}
\end{table}

\subsection{Data Quantity \& Type}
\begin{figure} 
\centering
\includegraphics[width=0.415 \paperwidth]{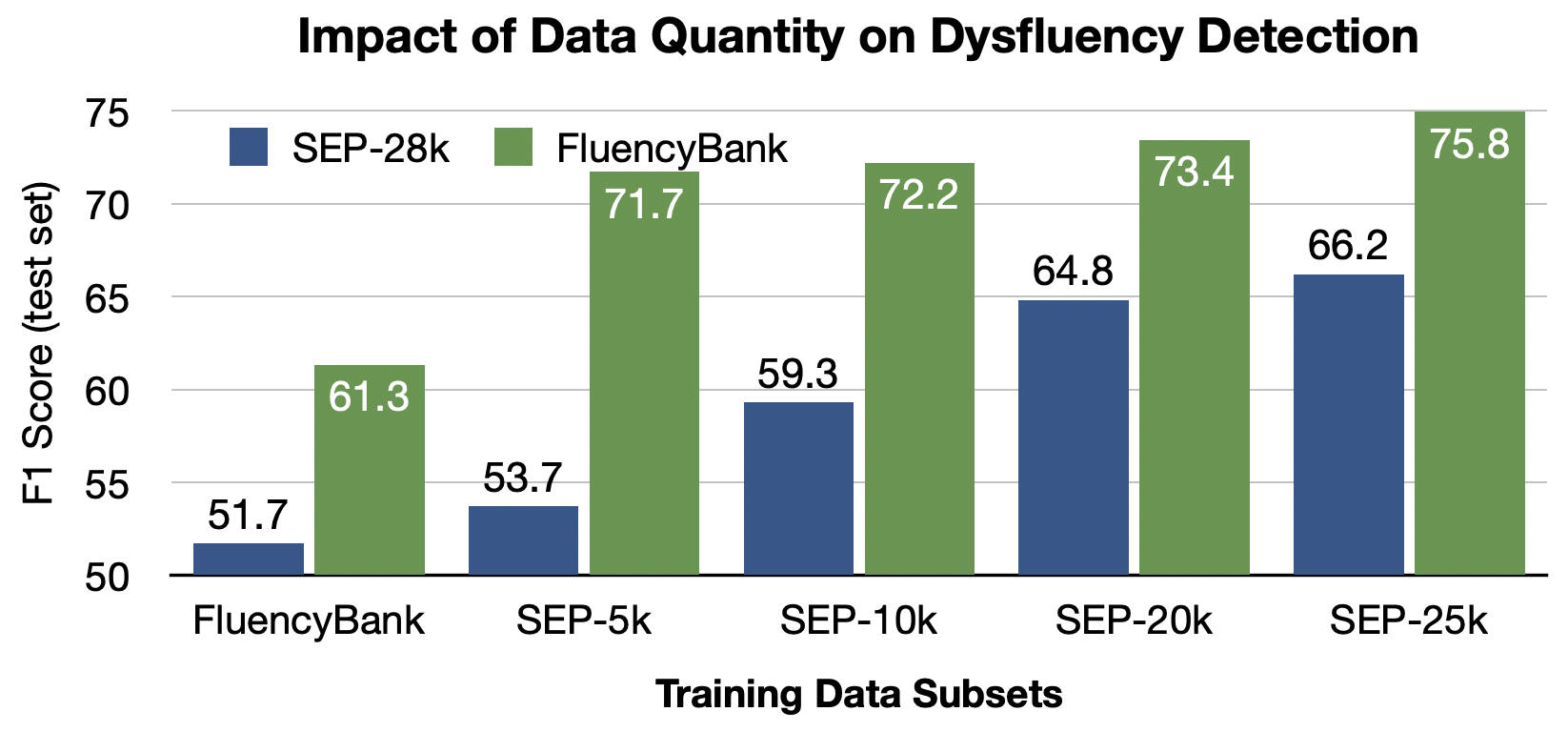}
\caption{Test performance when training models only on FluencyBank clips or subsets of clips from SEP-28k.}
\label{fig:ablation}
\end{figure}

The central hypothesis for this work was that existing datasets are too small and contain too few participants for training effective dysfluency detection models.
This is corroborated by results in Figure~\ref{fig:ablation} which shows performance on SEP-28k and FluencyBank while training on different subsets.
In the best case, there is a 24\% relative F1 improvement in FluencyBank when training on all 25k SEP training samples compared to the 3k FluencyBank set. 
Even using only 5k SEP clips already performs FluencyBank performance by 16\% F1.
This could be because there are a larger number of users in the dataset and the data contains more variability in speaking styles.
As expected, performance on SEP-28k is worst when training on FluencyBank and increases with larger numbers of training samples. 


\section{Conclusion}
We introduced SEP-28, which contains over an order of magnitude more annotations than existing public datasets and added new annotations to FluencyBank. 
These annotations can be used for many tasks so we encourage others to explore the data, labels, and splits in ways beyond what was is described here. 
Future work should explore alternative approaches, e.g., using language models, which may improve performance for some dysfluency types that are more difficult to detect.
Lastly, while dysfluencies are most common in those who stutter, future work should address how they can be detected from people with other speech disorders, such as dysarthria, which may be characterized differently.
\vspace{6pt}

\noindent \textbf{Acknowledgment}: Thanks to Lauren Tooley for countless discussions on the clinical aspects of stuttering.

\vfill\pagebreak

\bibliographystyle{IEEEbib}
\bibliography{strings,refs}

\end{document}